\documentclass[twocolumn]{aastex63}

\def\a{\alpha}
\def\b{\beta}
\def\g{\gamma}
\def\y{\gamma}
\def\d{\delta}
\def\D{\Delta}
\def\e{\epsilon}
\def\i{\iota}
\def\k{\kappa}
\def\l{\lambda}
\def\u{\mu}
\def\v{\nu}
\def\s{\sigma}
\def\t{\tau}

\def\x{\xi}

\def\n{\nabla}
\def\w{\omega}
\def\z{\zeta}

\def\p{\phi}

\def\L{\Lambda}

\def\O{\Omega}

\shorttitle{\textsc{Microlensing Events in the Galactic Plane}}
\shortauthors{Rodriguez \& Mróz}
\graphicspath{{./}{figures/}}

\usepackage{float}
\usepackage{appendix}
\usepackage{xcolor}
\usepackage{amsmath}
\usepackage{lipsum}

\begin{document}

\title{Microlensing Events in the Galactic Plane Using the Zwicky Transient Facility}

\correspondingauthor{Antonio C. Rodriguez}
\email{acrodrig@caltech.edu}

\author[0000-0003-4189-9668]{Antonio C. Rodriguez}

\affiliation{California Institute of Technology, Department of Astronomy, 
1200 E. California Blvd, 
Pasadena, CA, 91125, USA}

\author[0000-0001-7016-1692]{Przemek Mróz}

\affiliation{Astronomical Observatory, University of Warsaw, 
Al. Ujazdowskie 4, 00-478, 
Warszawa, Poland}
\affiliation{California Institute of Technology, Department of Astronomy, 
1200 E. California Blvd, 
Pasadena, CA, 91125, USA}

\author[0000-0001-5390-8563]{Shrinivas R. Kulkarni}
\affiliation{California Institute of Technology, Department of Astronomy, 
1200 E. California Blvd, 
Pasadena, CA, 91125, USA}

\author[0000-0002-8977-1498]{Igor Andreoni}
\thanks{Gehrels Fellow}
\affiliation{Joint Space-Science Institute, University of Maryland, College Park, MD 20742, USA}
\affiliation{Department of Astronomy, University of Maryland, College Park, MD 20742, USA.}
\affiliation{Astrophysics Science Division, NASA Goddard Space Flight Center, Mail Code 661, Greenbelt, MD 20771, USA.}
\author[0000-0001-8018-5348]{Eric C. Bellm}
\affiliation{DIRAC Institute, Department of Astronomy, University of Washington, 3910 15th Avenue NE, Seattle, WA 98195, USA}

\author[0000-0002-5884-7867]{Richard Dekany}
\affiliation{Caltech Optical Observatories, California Institute of Technology, 1200 E. California
             Blvd, Pasadena, CA 91125, USA}

\author{Andrew J. Drake}
\affiliation{California Institute of Technology, Department of Astronomy, 
1200 E. California Blvd, 
Pasadena, CA, 91125, USA}

\author[0000-0001-5060-8733]{Dmitry A. Duev}
\affiliation{California Institute of Technology, Department of Astronomy, 
1200 E. California Blvd, 
Pasadena, CA, 91125, USA}

\author[0000-0002-8532-9395]{Frank J. Masci}
\affiliation{IPAC, California Institute of Technology, 1200 E. California Blvd, Pasadena, CA 91125, USA}

\author{Thomas A. Prince}
\affiliation{Division of Physics, Mathematics, and Astronomy, California Institute of Technology, Pasadena, CA 91125, USA}
             
\author[0000-0002-0387-370X]{Reed Riddle}
\affiliation{California Institute of Technology, Department of Astronomy, 
1200 E. California Blvd, Pasadena, CA, 91125, USA}

\author[0000-0003-4401-0430]{David L. Shupe}
\affiliation{IPAC, California Institute of Technology, 1200 E. California Blvd, Pasadena, CA 91125, USA}

\begin{abstract}
Microlensing is a powerful technique to study the Galactic population of ``dark'' objects such as exoplanets both bound and unbound, brown dwarfs, low-luminosity stars, old white dwarfs, neutron stars, and almost the only way to study isolated stellar-mass black holes. The majority of previous efforts to search for gravitational microlensing events have concentrated towards high-density fields such as the Galactic bulge. Microlensing events in the Galactic plane have the advantage of closer proximity and better constrained relative proper motions, leading to better constrained lens mass estimates at the expense of a lower optical depth compared to events towards the Galactic bulge. We use the Zwicky Transient Facility (ZTF) Data Release 5 (DR5) compiled from 2018--2021 to survey the Galactic plane in the region of $|b| < 20^\circ$. We find a total of 60 candidate microlensing events including three that show a strong microlensing parallax effect. The rate of events traces Galactic structure, decreasing exponentially as a function Galactic longitude with scale length $\ell_0 \sim 37^\circ$. On average, we find Einstein timescales of our microlensing events to be about three times as long ($\sim60$ days) compared to those towards the Galactic bulge ($\sim20$ days). This pilot project demonstrates that microlensing towards the Galactic plane shows strong promise for characterization of dark objects within the Galatic disk.

\end{abstract}

\section{Introduction}
In a gravitational microlensing event, the light from a background source is magnified by a foreground lensing object as the light rays emitted from the source are deflected by the gravitational field of the lensing object \citep{paczynski86, paczynski96}. The lens, source, and observer are moving with respect to one another, and so a microlensing event can lead to a transient brightening and dimming of the source. Since microlensing is independent of the lens object brightness, it serves as one of the very few ways to study ``dark" objects such as exoplanets (rogue or otherwise), brown dwarfs, low-luminosity stars, neutron stars, and stellar mass black holes. Microlensing is almost the \textit{only} way to study isolated black holes. Probing the most massive of these stellar dark objects explores the most extreme densities and gravitational fields in the Universe, placing these studies at the forefront of fundamental physics.

However, from a qualitative description alone, it is clear that microlensing events have low probability to be observed from the vantage point of an astronomer on Earth. Thus, early dedicated microlensing surveys were directed towards the parts of the sky with the highest density of sources such as the Galactic bulge and the Large and Small Magellanic Clouds. These early surveys included Expérience pour la Recherche d'Objets Sombres \citep[EROS;][]{eros_93}, Optical Gravitational Lensing Experiment \citep[OGLE;][]{udalski92}, the search for MAssive Compact Halo Objects \citep[MACHO;][]{macho_1993}, and Microlensing Observations in Astrophysics \citep[MOA;][]{moa_2008}. Early surveys purposefully avoided the more sparsely populated Galactic plane, where finding microlensing events was deemed to be difficult due to the highly reduced amount of both sources and lenses, and would thus lead to very low event rates per unit area on the sky. 

The Zwicky Transient Facility (ZTF) is undertaking a survey of the entire Northern sky down to $\sim 21$ mag making use of a 47 $\textrm{deg}^2$ field-of-view camera, operating on a 1 to 3-day cadence. A several-day cadence is sufficient for detecting stellar microlensing events in the Galactic bulge \citep{2012gaudi_review}, and even better for microlensing events in the Galactic plane, where they are expected to have longer timescales \citep{sajadian2019}. \cite{gould2013} strongly advocated for a microlensing survey of the Galactic plane as these events are much more likely to be longer, which allows for the measurement of the ``microlensing parallax'' from the lightcurve. The microlensing parallax (not to be confused with the traditional distance parallax) causes deviations from the typical microlensing lightcurve due to the projection of the lens onto Earth's revolution around the Sun. The microlensing parallax provides another constraint on the lens mass by allowing an observer on Earth to view the microlensing event from two different vantage points. 

Better yet, in the Galactic plane, microlensing events are also more likely to be located closer to Earth, which makes the measurement of the microlensing astrometric effect easier \citep[e.g.][]{rybicki2018}. Astrometry provides another constraint on the mass of the lens, leading to a possible total of three independent mass constraints that are more difficult to obtain for Galactic bulge events. Thus, a microlensing survey of the Galactic plane would reveal the distribution of exoplanets, isolated brown dwarfs, neutron stars, and black holes, as well as provide a novel exploration the structure of the disk of the Milky Way. Theoretical predictions of the detection of Galactic plane microlensing events by the Rubin Observatory Legacy Survey of Space and Time (LSST) were undertaken by \cite{sajadian2019} and \cite{street2018}, and by ZTF includes \cite{medford2020}.

EROS, as part of a three-year survey, looked at small regions in the sky towards the Galactic spiral arms \citep{eros_plane2001}. This study revealed the lower overall optical depth to microlensing for events located towards the plane compared to events located towards the bulge. The OGLE Galaxy Variability Survey \citep[GVS;][]{udalski2015} looked towards the Galactic equator in the range $0^\circ < 
\ell < 50^\circ$ and $190^\circ < \ell < 360^\circ$, and found that microlensing events in the plane are three times longer than events found towards the Galactic bulge \citep{mroz2020_ogle_plane}. Both EROS and OGLE studies were concentrated in the southern hemisphere, while our study is located in the northern hemisphere. A pilot search through ZTF Data Release 2 (DR2) data already revealed a sizeable population of microlensing candidate events \citep{2020mroz_galacticplane}, paving the way for a bigger study.

In this work, we search the three-year-long archive of Zwicky Transient Facility (ZTF) Data Release 5 (DR5) to find microlensing events in the Galactic plane. In Section \ref{sec:data}, we describe our data collection process. In Section \ref{sec:event_selection}, we outline our methodology including our accounting for false positives. In Section \ref{sec:results}, we present the candidate microlensing events along with model parameters. We conclude with a discussion of the detection efficiency of our algorithm, and discuss implications on Galactic structure.

\section{Data}
\label{sec:data}
The Zwicky Transient Facility (ZTF) is a photometric survey that uses a wide 47 $\textrm{deg}^2$ field-of-view camera on the Samuel Oschin 48-inch telescope at Palomar Observatory with $g$, $r$, and $i$ filters \citep{bellm2019, graham2019, dekanyztf, masci_ztf}. In its first year of operations, ZTF carried out a public nightly Galactic Plane Survey in $g$-band and $r$-band \citep{ztf_northernskysurvey_bellm, kupfer_ztf}. This survey is in addition to the Northern Sky Survey which operates on a 3 day cadence \citep{bellm2019}. The pixel size of the ZTF camera 1" and the median delivered image quality is 2.0" at FWHM. 

We use of ZTF Data Release 5 (DR5), and analyze all available Galactic plane fields ($|b| < 20^\circ$), which are a combination of public data taken from March 2018--January 2021 as well as the ZTF partnership and Caltech-exclusive data up to April 2021. Lightcurves have a photometric precision of 0.01 mag at 13--14 mag down to precision of 0.1--0.2 mag for the faintest objects at 20--21 mag. Due to Galactic reddening, we begin by only analyzing the brighter $r$-band lightcurves.

The entire data download process is done using Kowalski, a data archive tool developed for the ZTF collaboration\footnote{\url{https://github.com/dmitryduev/kowalski}} \citep{duev2019}. Potential events are followed up with baseline corrected forced photometry provided for the ZTF collaboration through the Infrared Processing and Analysis Center\footnote{\url{https://irsa.ipac.caltech.edu/Missions/ztf.html}} \citep[IPAC;][]{masci_ztf}. While both raw photometry and forced photometry are PSF-fit photometry, the forced photometry calculates photometry of the object on all available frames by forcing the location of the PSF to remain fixed according to the ZTF absolute astrometric reference. This has the effect of reducing the uncertainties compared to the raw photometry since ZTF absolute astrometry for frames is generally better than individual point source astrometry. All final models are based on data taken from forced photometry. In total, we analyzed 194 fields, and the resulting trove has $\sim 2 \times 10^9$ sources which amounts to 535,218,421 lightcurves that contain at least 30 data points.

\section{Event Selection}
\label{sec:event_selection}
In essence, a microlensing event is an apparent amplification of an otherwise steady source. The simplest model of a microlensing lightcurve is the point-source, point-lens model (PSPL). The model is described by the following equation:
\begin{gather}
    F(t) = \begin{cases}
        F_s A(t; t_0, u_0, t_\mathrm{E}) + F_b & \textrm{ no parallax}\\
        F_s A(t; t_0, u_0, t_\mathrm{E}, \pi_\textrm{EN}, \pi_\textrm{EE}) + F_b & \textrm{ with parallax}
    \end{cases}
    \label{eq:flux}
\end{gather}
where $F_s$ is the source flux and $F_b$ the blended flux in the aperture that is not magnified during the lensing event. $A(t)$ is the magnification as a function of time. $t_0$ is time of closest approach between the source and the lens, $u_0$ is the effective impact parameter at the time of closest approach, and $t_\mathrm{E}$ is the Einstein timescale of the event. The microlensing parallax vector, $\boldsymbol{\pi}_\textrm{E}$, is split into its North and East components, $\pi_\textrm{EN}, \pi_\textrm{EE}$, which characterizes the effect of the Earth's orbital rotation around the Sun (see Section \ref{sec:parallax}). We refer the reader to reviews such as \cite{paczynski96} and \cite{2012gaudi_review} for a detailed derivation of the lightcurve shape and discussion of model parameters. Modeling of PSPL events and events featuring microlensing parallax are implemented efficiently in \texttt{MulensModel}\footnote{\url{https://rpoleski.github.io/MulensModel/index.html}}\citep{mulensmodel2019}, which we use for all modeling in this study.

\subsection{Methodology}
\label{sec:selection_methodology}
We perform the series of cuts shown in Table \ref{tab:data_cuts} to find microlensing events in ZTF data. We begin by inspecting raw photometry lightcurves in the $r$-band. While even less affected by reddening, we did not use $i$-band data from ZTF due to its substantially lower number of epochs which are unsuitable for finding microlensing events. We inspect $g$-band data, if available, when obtaining forced photometry. All cuts were initially tested to rediscover the maximum number of previously found events in \cite{2020mroz_galacticplane}, and were iteratively tested on simulated events injected into ZTF lightcurves and updated. Not all events reported in \cite{2020mroz_galacticplane} are rediscovered due to some being binary events with lightcurves strongly departing from the PSPL model. By limiting our events to those with $t_0 \pm 0.3t_\mathrm{E}$ taking place within the ZTF DR5 observing baseline, we exclude some previously reported candidate events. This is done to ensure a more pure sample at the end of our selection process. Given our model assumptions and quality cuts, we recover $\sim75$\% of the events reported in \cite{2020mroz_galacticplane}. 

\subsubsection{Skewness-Von Neumann Space}\label{sec:gamma_eta_discussion}
We begin with $r$-band lightcurves that each contain at least 30 data points. The first cut we apply to our data set serves two purposes: 1) pick out statistically significant deviations during quiescence and 2) find brightening as opposed to dimming events in an otherwise static lightcurve. The Von Neumann statistic, $\eta$, is meant to do the former, and the skewness, $\gamma$, the latter:
\begin{gather}
     \eta \textrm{ (von Neumann) } = \sum_{i=2}^N\frac{(x_i - x_{i-1})^2}{(N-1)\hat{\sigma}^2}\\\
     \gamma \textrm{ (skewness) } = \sum_{i=1}^N\frac{(x_i - \overline{x})^3}{N\hat{\sigma}^3}
\end{gather}
where $\hat{\sigma}$ is the sample standard deviation. The von Neumann statistic is defined as the mean square successive difference divided by the sample variance. Therefore, statistically significant bursts will lead to an overall large variance, while keeping a small successive difference that make $\eta$ small in those events. So we are looking for small $\eta$. Furthermore, a positive skewness corresponds to a dimming event, while a negative skewness corresponds to a brightening event. \cite{2016wyrzykowski} found an explicit condition similar to the one outlined in Table \ref{tab:data_cuts} in $\gamma$--$\eta$ space, while we modify it to find all events previously found in \cite{2020mroz_galacticplane}. Our condition in $\gamma$-$\eta$ space is $\gamma < 0.9 \textrm{ and } \log_{10}(0.9-\gamma) > -1.25\log_{10}(1/\eta) + 0.75$.

\subsubsection{Cuts Based on Modeling}
As seen in Table \ref{tab:data_cuts}, the initial cut on all data based on skewness-von Neumann space reduces the sample size by a factor of $\sim 3\times 10^{-4}$. We then fit the data to the PSPL model. In our modeling routine, we account for possible outliers in the data due to anomalous effects such as poor weather or corrupted data. We pick a fixed ``lensing window'' of 360 days during which a lensing event may take place \citep[following the methodology of][]{mroz2017nature}. We remove lightcurves for which fewer than 6 magnified points are located within this window, as well as lightcurves for which the remaining number of points is fewer than 30. These cuts ensure that we have a sufficient amount of data points for modeling a microlensing event within the lensing window as well as enough data points outside the window the determine. The latter is done to ensure the source was quiescent outside of the lensing window and discard the possibility of it being an intrinsically variable source.

We also remove events with an outburst magnitude (mag difference between peak of microlensing model and quiescent state) less than 0.2 mag. This removes false positives such as outbursting Be stars or anomalous weather conditions leading to photometric variations. We then calculate the reduced chi-squared statistic:
\begin{gather}
    \chi^2_\textrm{red} = \frac{1}{N-m}\sum_i^N \frac{\left(F_\textrm{model}(t_i) - F_\textrm{data}(t_i)\right)^2}{\sigma_i^2}
\end{gather}
where $N$ is the total number of data points, $m$ is the number of the parameters in the fitted model, and $F_\textrm{model}$ and $F_\textrm{data}$ are the best-fit model flux and observed flux, respectively. We discard lightcurves for which $\chi^2_\textrm{red} < 3$ both over the lensing window as well as over the entire lightcurve. We calculate this statistic for both cases to ensure to avoid false positives where a non-quiescent source undergoes a well-modeled outburst or situations where a non-variable source undergoes a poorly-modeled outburst.

We then exclude events for which $t_0 \pm 0.3t_\mathrm{E}$ is not fully located within the available data. This ensures that we sample enough of the lightcurve to assess its symmetry about $t_0$ and ensure it is a microlensing event and not an eruptive phenomenon with an asymmetric outburst. Models with $u_0 > 1$ would require amplitudes lower than $\sim 0.3$ mag, which are difficult to disentangle from noise or false positives within our sample. Finally, we place cuts based on the best-fit characteristic timescale, $t_\mathrm{E}$. Since the overall temporal baseline for ZTF DR5 is $\sim1100$ days, we exclude events with $t_\mathrm{E} > 500
$ days. We also exclude events with amplitudes less than 0.3 mag if they are longer than 80 days. This cut filters our long-period, low-amplitude variables.

After all these cuts, we are left with $\sim 1 \times 10^{-6}$ of our original sample (see Table \ref{tab:data_cuts}). We visually inspect the $r$-band raw photometry for false positives. We then request IPAC forced photometry of the remaining candidates in $r$-, $g$- and $i$- bands. Note that $i$-band photometry is only used to discard false positives such as long period variable stars. Using both $r$- and $g$-band photometry, we exclude events that are not achromatic such as supernovae and variable stars. While iterating this process over different series of cuts, we serendipitously found another microlensing event and included it as part of our final sample of 60 events -- a sample size $\sim 10^{-7}$ of the initial set of lightcurves in ZTF DR5.

\begin{deluxetable*}{ccc}
\label{tab:data_cuts}
\tabletypesize{\small}

\tablehead{
\colhead{Criteria}  & \colhead{Remarks} & \colhead{Number}
}
\tablecaption{Cuts on ZTF Data to Find Microlensing Events}
\startdata
Begin with all objects in ZTF Galactic plane fields & Must have at least 30 data points. & 535,218,421\\
\hline
$\gamma < 0.9 \textrm{ and } \log_{10}(0.9-\gamma) > -1.25\log_{10}(1/\eta) + 0.75$ & All statistically significant outbursts. & 174,950\\
Model Converges & PSPL model converges. Removes flat lightcurves. & 108,116\\
\hline
Algorithm tested on prev. found events and simulated data: & &\\
$N_\textrm{pts, lens} \geq 6$ & Significant number of magnified points within lens window. & \\
$\textrm{Amplitude} \geq 0.2$ & Small-amplitude false positives removed. & \\
$\chi^2_\textrm{red, tot} \leq 3$ & Full lightcurve modeled well. & \\
$\chi^2_\textrm{red, lens} \leq 3$ & Lens window modeled well. & \\
$t_0 \pm 0.3t_\mathrm{E}$ fully in data & Removes events before/after DR5 baseline. & \\
$u_0 \leq 1$ & Only sensible values of $u_0$ kept. & \\
$0 \leq t_\mathrm{E}  \leq 500$ days & Long timescale contaminants removed. & \\
$\textrm{Amplitude} \geq 0.3 \textrm{ if } t_\mathrm{E} \geq 80$ days & Low-amplitude, long-period variables removed. & 606\\
\hline
Visual Inspection & All candidates scanned with raw photometry & 92\\
With Forced Photometry & Reliable photometry removes final contaminants & 59\\
Re-scan & Visual rescanning with different cuts & 60
\enddata
\caption{All cuts applied to ZTF DR5 data to find microlensing events. The final algorithm finds 59 events, and another was serendipitously found after testing different cuts on the data, leading to a total of 60 events.}
\end{deluxetable*}

\subsection{MCMC Parameter Exploration}
We use Markov Chain Monte Carlo (MCMC) through the \texttt{emcee} package \citep{hogg2018} as a way to obtain quantitative estimates of model parameters and their uncertainties for each of our sample microlensing events. Microlensing lightcurves exhibit a strong continuous degeneracy between $t_\mathrm{E}$ and $u_0$, where a large Einstein timescale can be compensated by decreasing the value of the impact parameter. There are also discrete degeneracies when parallax parameters, $\pi_\textrm{EN}$ and $\pi_\textrm{EE}$ are introduced, which can then lead to two or four possible models.

Since MCMC provides a natural way to provide physical priors on our model as well as quantitatively explore parameter degeneracies, we use it to provide model parameters for each event once forced photometry is obtained. We set flat priors on $t_\textrm{eff} = u_0t_\mathrm{E}$ between 0 and 2000 days, and a flat prior on $t_0$ and for $u_0 \geq 0$ (note that we allow $u_0 < 0$ in parallax models). If both $r$- and $g$-band data are available, we fit all data simultaneously. Since $F(t)$ is linear in both $F_s$ and $F_b$ (Equation \ref{eq:flux}), we need not include them in the likelihood function of the MCMC analysis and can simply extract their posterior values from the other model parameters. We do, however, include them in the prior to allow for blending:
\begin{gather}
    \mathcal{L}_\textrm{prior} = 
    \begin{cases}
    1 & \textrm{ if } F_b \geq 0\\
    \exp\left(-\frac{F_b^2}{2\sigma^2}\right) & \textrm{ if } F_b < 0
    \end{cases}
\end{gather}
where $\sigma = F_\textrm{min}/3$ and $F_\textrm{min}$ is the flux corresponding to $r = 21$. In Section \ref{sec:results}, we report credible intervals between the 16- and 84- percentiles for parameters $u_0, t_0, t_\mathrm{E}$, and also $\pi_\textrm{EE}, \pi_\textrm{EN}$ if we find the parallax effect to be statistically significant (see Section \ref{sec:parallax}).

\subsection{Events with Microlensing Parallax}
\label{sec:parallax}

The microlensing parallax is an effect that describes the projection of the effective angular radius of the lens (known as the angular Einstein radius) onto Earth as it orbits the Sun, and introduces small deviations from the typical PSPL lightcurve \citep{formalism_gould2000}.

In our search algorithm, we pick out events with statistically significant signatures of the parallax effect once we obtain forced photometry lightcurves in $r$- and $g$-band data, if available. The algorithm starts with the final, visually confirmed sample of microlensing events. We then fit models allowing for parallax parameters $\pi_\textrm{EN}$ and $\pi_\textrm{EE}$. If $\chi^2_\textrm{parallax} <  \chi^2_\textrm{non-parallax} + \sigma^2$, where we empirically choose $\sigma = 4$, we declare it to be a possible parallax microlensing event. We now construct cumulative distribution functions (CDFs) for the quantity $\chi^2_\textrm{non-parallax} - \chi^2_\textrm{parallax}$ across the full baseline to visually determine which events show a steady contribution to the CDF across the lensing window. We discard events for which the fit improves mostly over the baseline, or mostly over just one half of the lensing window. The resulting number of events from the purely statistical cut leaves 13 possible parallax events, and when combined with the CDF visual analysis, we are left with 3 events that show compelling evidence of parallax.

\section{Results and Discussion}
\label{sec:results}
\begin{figure*}
    \centering
    \includegraphics[ scale=0.43]{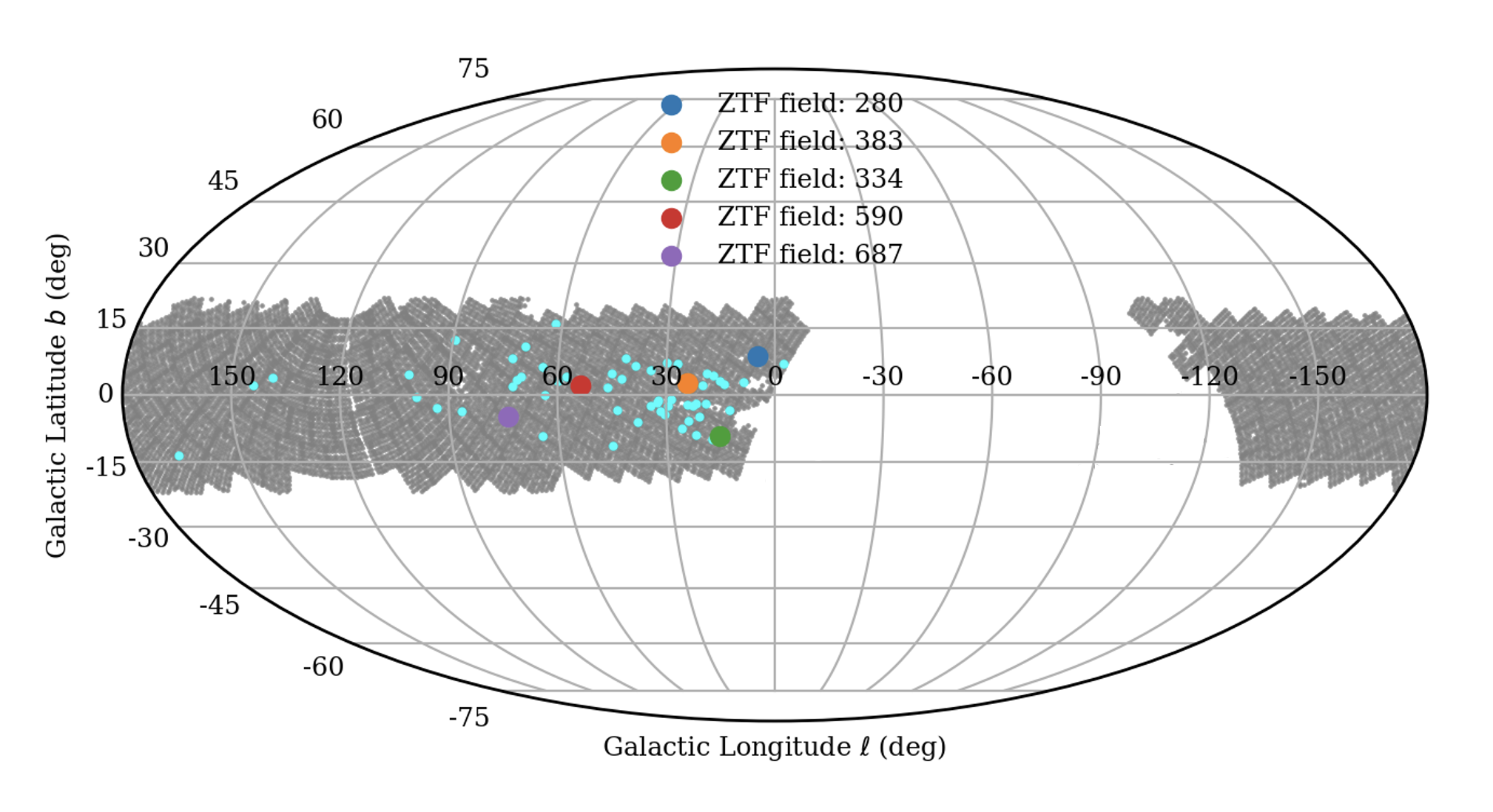}
    \caption{Location of all 60 microlensing events found in ZTF data (cyan) along with highlighted fields shown in Figure \ref{fig:efficiencies}. The majority of all events are found within 10 degrees of the Galactic plane, despite looking 20 degrees in each extent in latitude. All scanned fields are shown in gray. More events are found towards the higher density lower Galactic longitudes.}
    \label{fig:fields_map}
\end{figure*}

Appendix Table \ref{tab:all_events} lists all 60 events found in ZTF DR5, along with model parameters resulting from the MCMC fit. We find that 22 of the 60 events were previously reported by \cite{2020mroz_galacticplane}, leading to 38 new events. The location of analyzed fields and event locations are presented in Figure \ref{fig:fields_map}. 

\subsection{Detection Efficiencies}
Once we obtain our final sample of 60 microlensing events, we identify all ZTF fields where events are located. In each field, we randomly select 100,000 indices corresponding to the observed lightcurve of a source within the field. Following the method of \cite{mroz2020_ogle_plane}, we then inject a simulated microlensing lightcurve taking place within the observing baseline of ZTF DR5, with parameters $u_0, t_0, \log t_\mathrm{E}$ selected from a flat distribution. We do not inject lightcurves with the microlensing parallax effect. 

\begin{figure}
    \centering
    \includegraphics[scale=0.42]{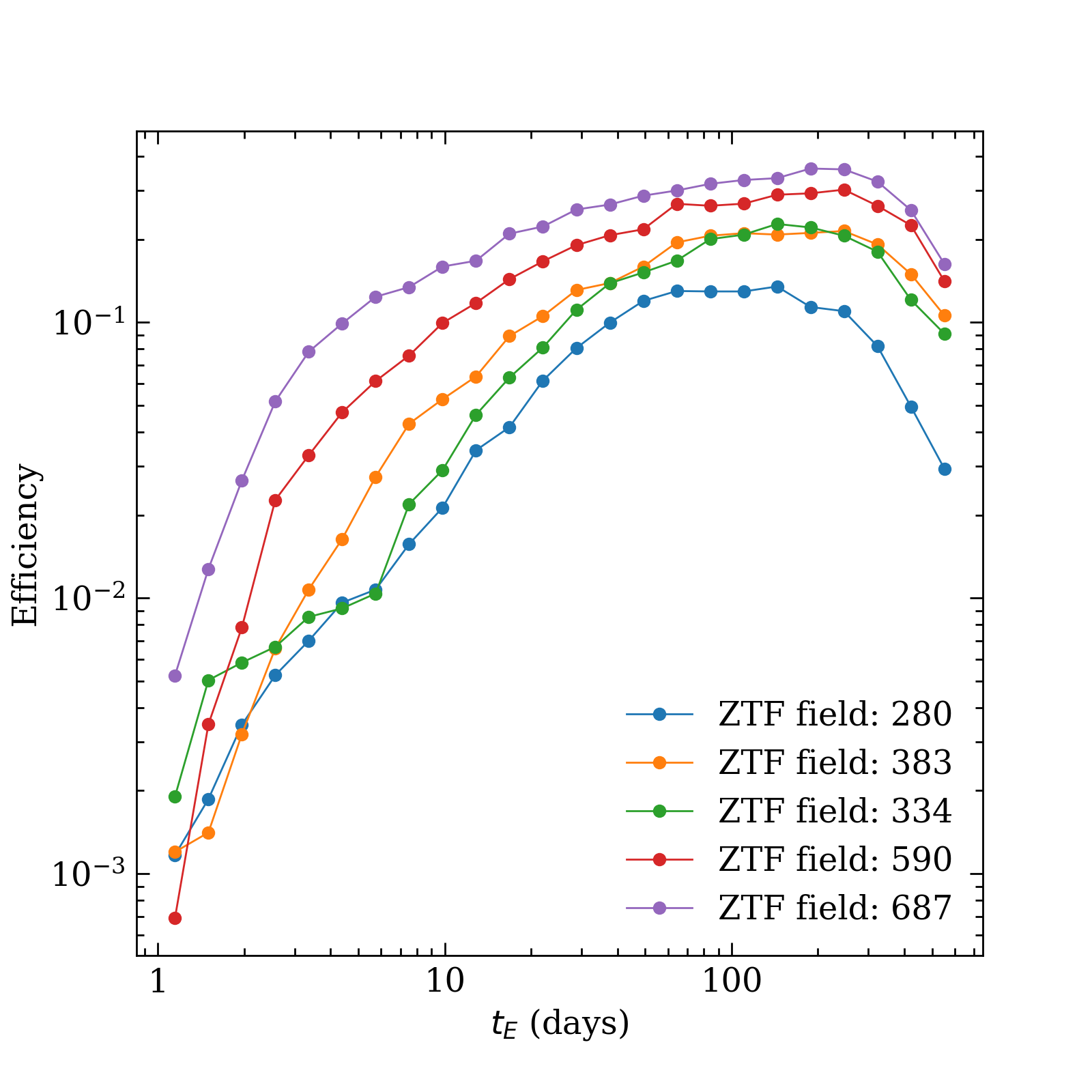}
    \caption{Detection efficiency for various ZTF fields where microlensing events are detected. Detection efficiency at all timescales decreases towards the Galactic Center. Independent of field, efficiency decreases towards small and large Einstein timescales.}
    \label{fig:efficiencies}
\end{figure}

We then run our event selection methodology as described in Section \ref{sec:selection_methodology} and see if the event extracted passes our series of cuts or not. After running this over all simulated events, over all ZTF fields where we have found one of our 60 sample events, we construct efficiency curves shown in Figure \ref{fig:efficiencies}. The efficiency is defined as the ratio of detected events to the injected events for bins in $t_\mathrm{E}$. Figure \ref{fig:fields_map} shows the location in Galactic coordinates where all 60 microlensing events are located, and additionally highlights the fields for which detection efficiencies are plotted in Figure \ref{fig:efficiencies}. 

In combination, those figures show that on the whole, detection efficiencies are lower for events located closer to the Galactic center. This is a combined effect of the reduced amount of time that these low-declination fields were observed by ZTF as well as non-optimal photometry in the crowded areas of this part of the Galaxy. Our simulations show that if a microlensing event takes place during a seasonal gap in observing (i.e. ZTF cannot observe every field of the sky continuously throughout the year), then our detection algorithm will fail to find it due to limited data.

Events with $t_\mathrm{E}$ between 1--10 days are hard to detect, as seen in Figure \ref{fig:efficiencies}, leading to a sharply dropping efficiency with timescale, while efficiencies remain mostly flat for $t_\mathrm{E}$ between 10--100 days. Longer timescale events with $2t_\mathrm{E}$ approaching the entire observing window of ZTF DR5 ($\sim 1100$ days) become hard to detect and lead to a decreasing efficiency at the largest timescales.

\subsection{Distribution of Einstein Timescales}
\begin{figure}
    \centering
    \includegraphics[scale=0.35]{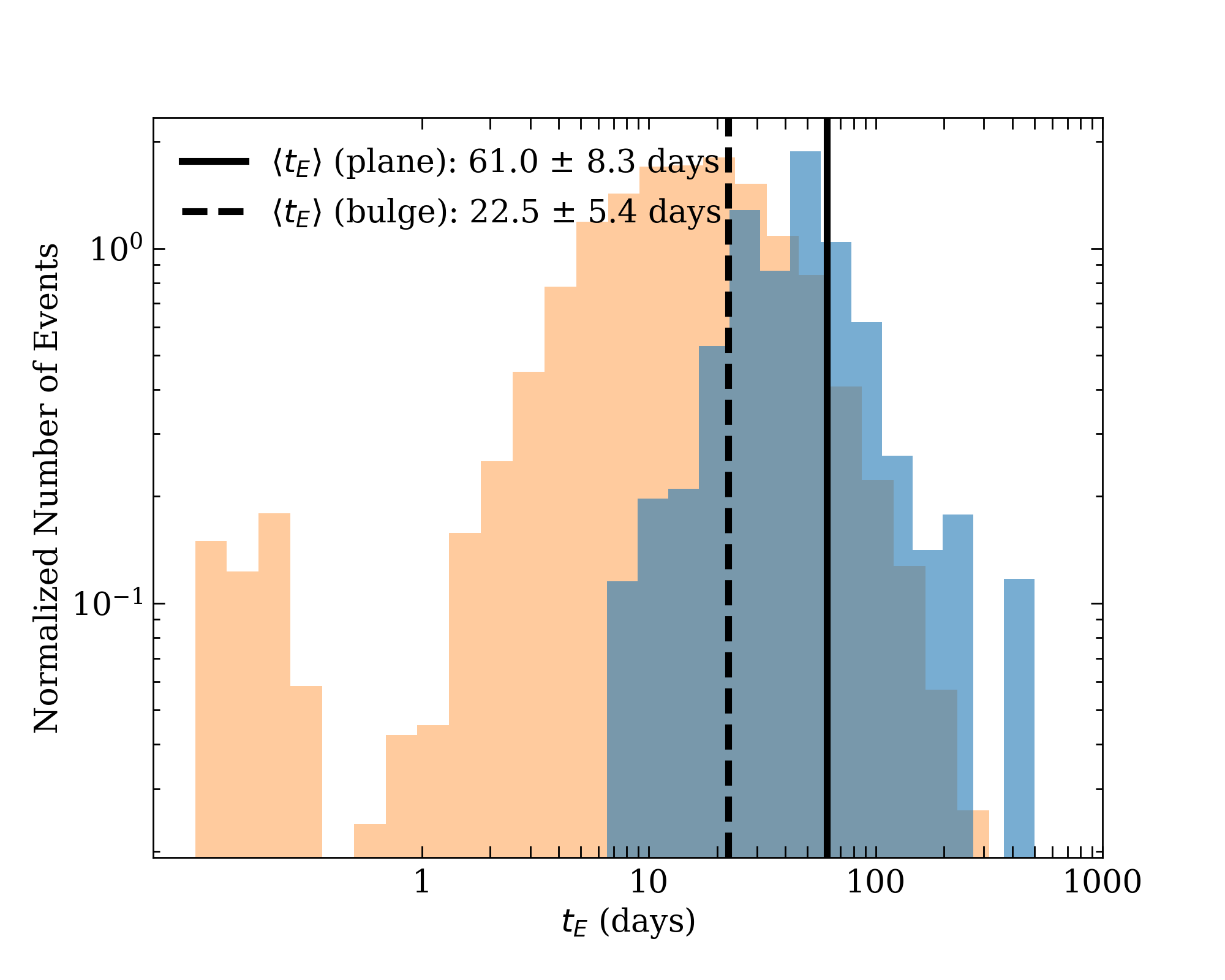}
    \caption{Binned Einstein timescales of all 60 events. Number densities are weighted by the reciprocal of the relevant efficiencies in order to obtain a statistically-corrected sample. In blue are results from this work: $\langle t_\textrm{E} \rangle \textrm{ (plane)} = 61.0 \pm 8.3$ days. In orange are results from \cite{mroz2017nature}, where $\langle t_\textrm{E} \rangle \textrm{ (bulge)} = 22.5 \pm 5.4$ days. Galactic plane microlensing events are, on average, three times longer than those in the bulge.}
    \label{fig:tE}
\end{figure}

Given detection efficiencies ($\varepsilon_i$ for field $i$), we calculate the mean Einstein timescale:
\begin{gather}
    \langle t_\mathrm{E} \rangle = \dfrac{\sum_i t_{E,i}/\varepsilon_i (t_{E,i})}{\sum_i 1/\varepsilon_i (t_{E,i})}
\end{gather}
\label{eq:avg_tE}
where each efficiency is evaluated at the relevant Einstein timescale. Figure \ref{fig:tE} shows the distribution of Einstein timescales, weighted by the reciprocal of detection efficiency to create a statistically-corrected histogram. We obtain a mean Einstein timescale of $\langle t_\mathrm{E} \rangle = 61.0 \pm 8.3$ days, demonstrating that on average, Galactic plane microlensing events are about three times are long as Galactic bulge events, which have $\langle t_\textrm{E} \rangle \textrm{ (bulge)} = 22.5 \pm 5.4$ days \citep{mroz2017nature}. This is in agreement with a study of the southern Galactic plane which found $\langle t_\mathrm{E} \rangle = 61.5 \pm 5.0$ days \cite{mroz2020_ogle_plane}.

\subsection{Distribution of Events and Galactic Structure}
Plotting the distribution of events as a function of Galactic longitude and latitude serves as an independent probe of Galactic structure. Figure \ref{fig:fields_map} demonstrates that while we searched Galactic fields up to $|b| < 20^\circ$, the majority of microlensing events are located within $10^\circ$ of the Galactic plane.

\begin{figure*}
    \centering
    \includegraphics[scale=0.36]{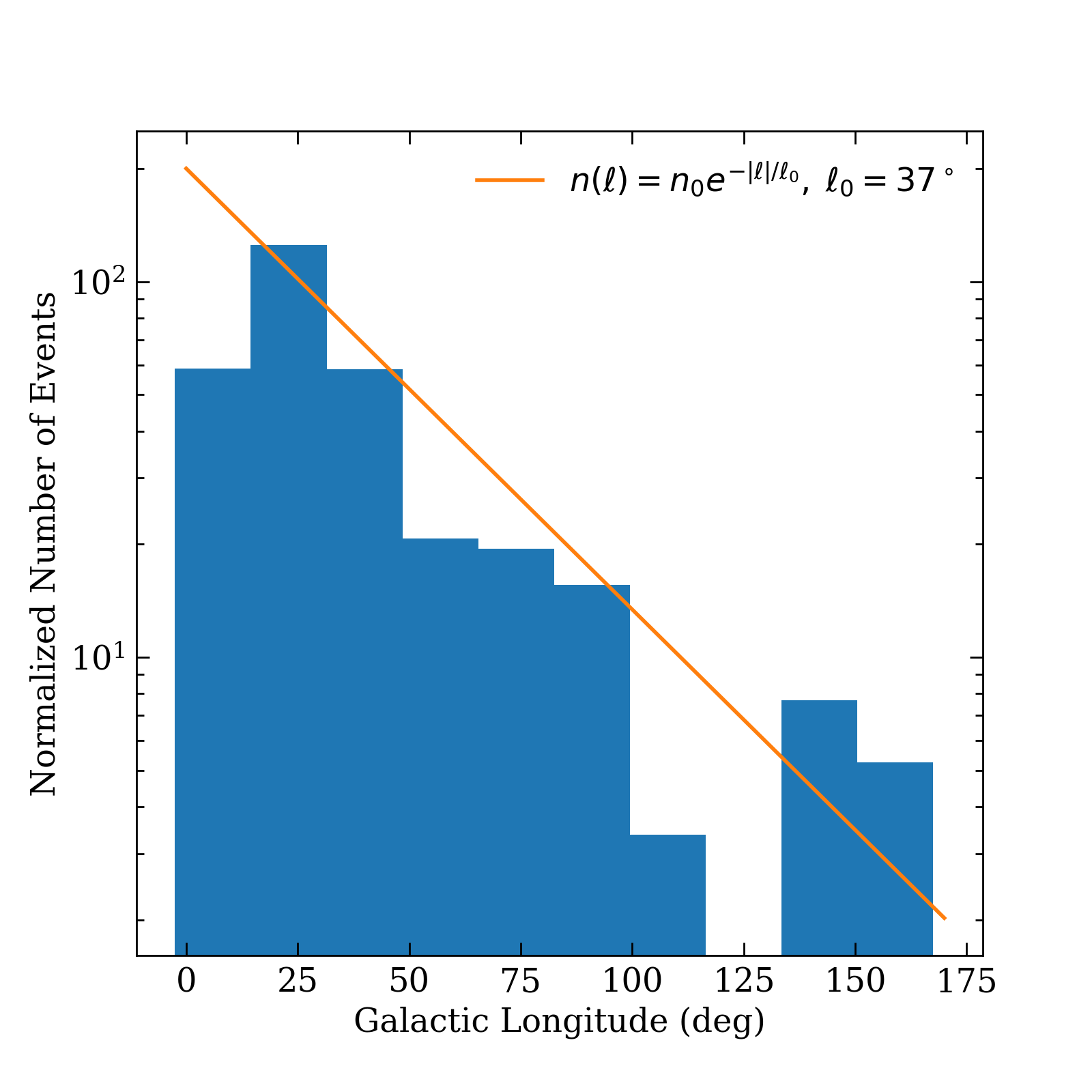}\includegraphics[scale=0.36]{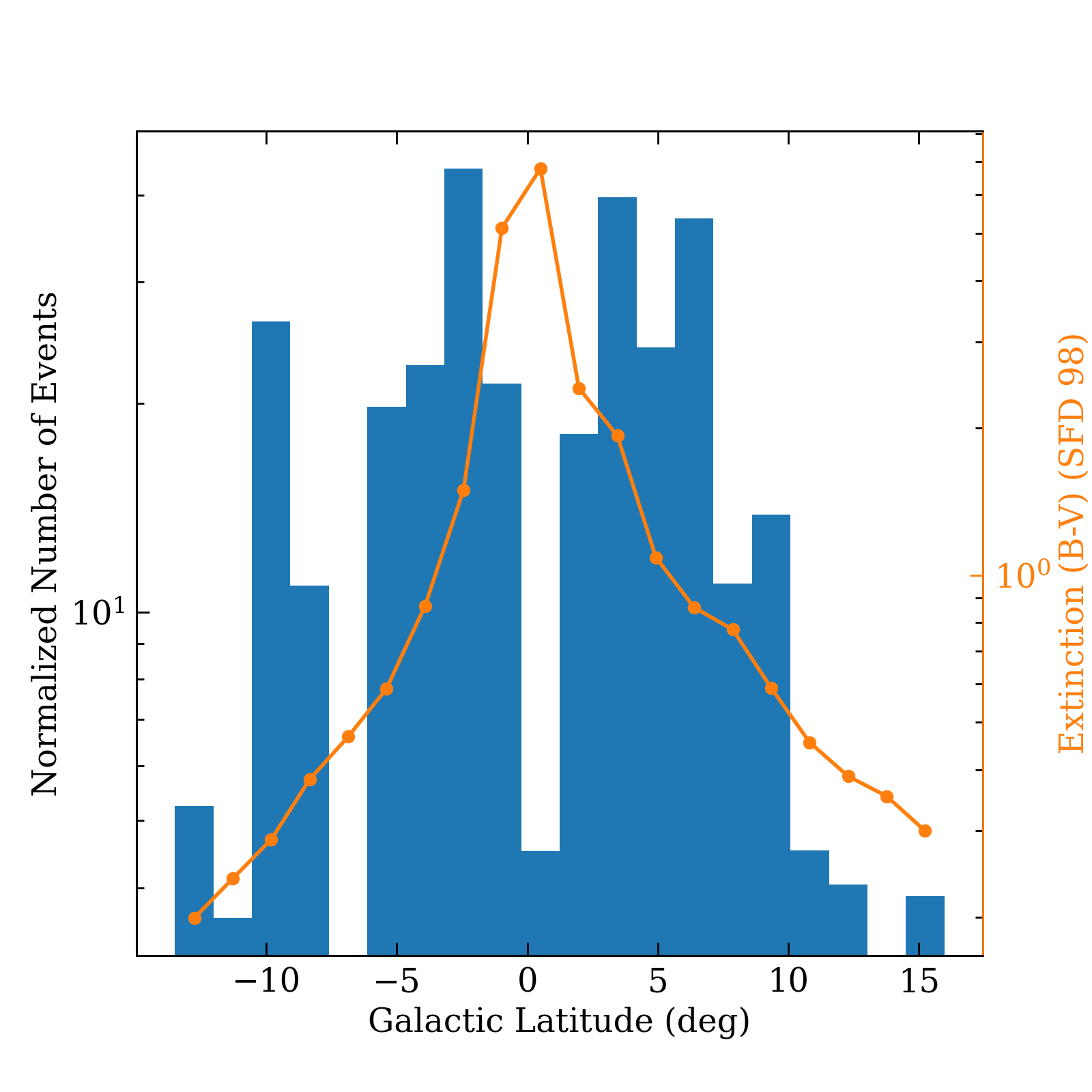}
    \caption{Number of events as a function of Galactic coordinates. Left: the number of events decreases exponentially with characteristic angular scale $\sim37^\circ$ as a function of longitude due to the decreased density of lensing objects and sources in the Galaxy at high longitudes. Right: the number of events decreases as a function of absolute latitude and  drops at $b=0^\circ$ due to extinction along the Galactic plane. Extinction model from \cite{sfd98} is adopted.}
    \label{fig:distribution_bins}
\end{figure*}

As can be seen from Figure \ref{fig:distribution_bins}, the number of events decreases with increasing (absolute) longitude and latitude, tracing the decreasing density of objects in the Galaxy in those regions. Assuming an exponential distribution of microlensing events as a function of absolute longitude, $n(\ell) = n_0 e^{-|\ell|/\ell_0}$, we use maximum likelihood estimation to find a characteristic angular scale of $\ell_0 = 37^\circ \pm 4^\circ$, which is consistent with that found by \cite{mroz2020_ogle_plane}. Moreover, the right panel of Figure \ref{fig:distribution_bins} shows that the normalized number of events found towards the Galactic midplane ($|b| \approx 0^\circ$) decreases dramatically due to strong interstellar reddening. To represent extinction, $E(B-V)$ is plotted, taken from \cite{sfd98}. This is in reasonable agreement with the results presented by \cite{mroz2020_ogle_plane}, but more detailed statistics, which would result from a larger dataset, are needed to confidently conclude this is the case in ZTF data. 

\begin{figure}
    \centering
    \includegraphics[scale=0.35]{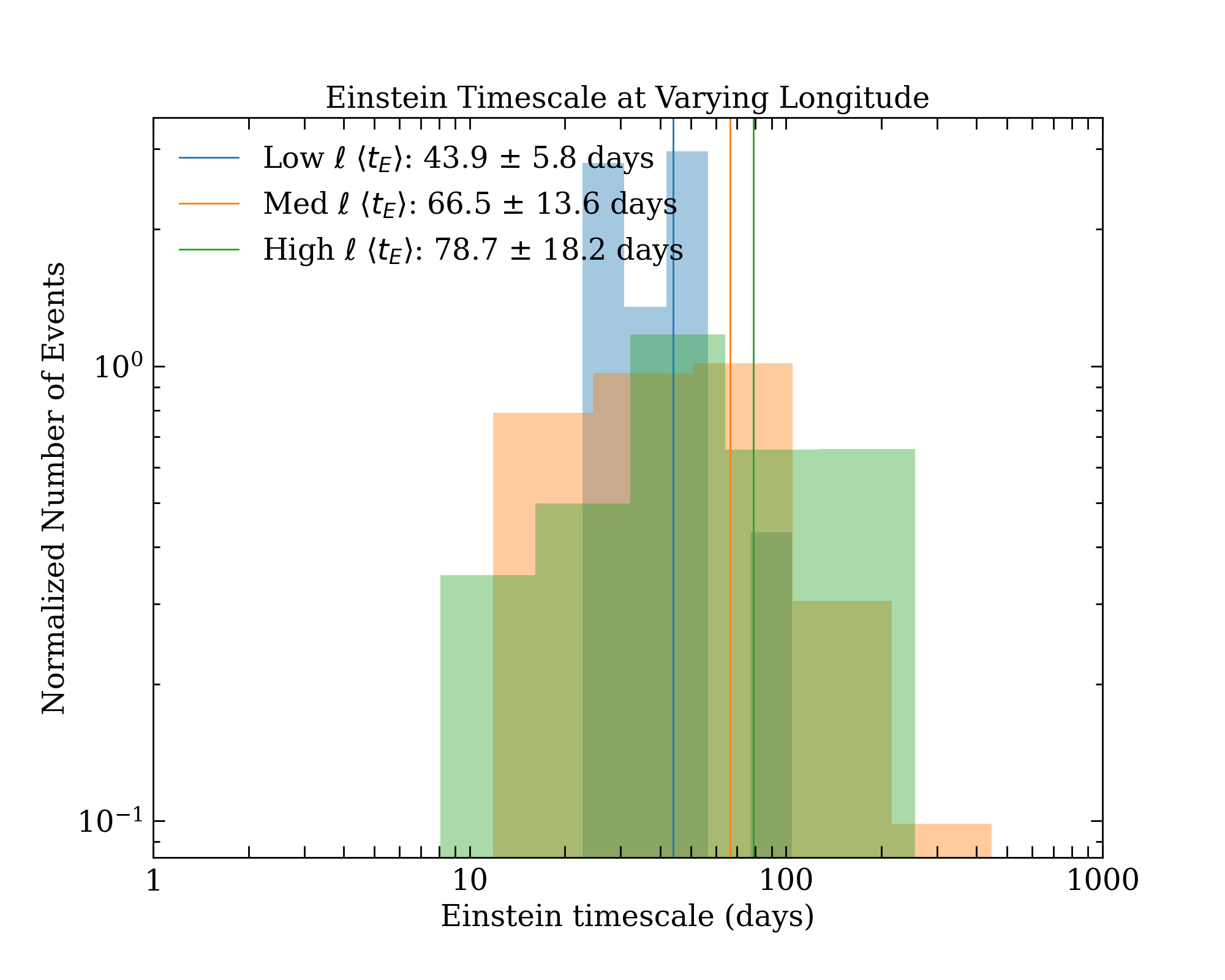}
    \caption{Number of events plotted as a function of Einstein timescale for three different longitude bins. Events at low longitude, $\ell < 20^\circ$, which trace regions closer to the Galactic bulge, are significantly shorter than those at $\ell > 20^\circ$, which trace the Galactic plane. There is no statistically significant trend between Galactic longitude and Einstein timescale for events in the Galactic plane beyond $\ell > 20^\circ$.}
    \label{fig:distribution_bins2}
\end{figure}

Figure \ref{fig:distribution_bins2} shows the number of events as a function of Einstein timescale, but in three separate longitude bins: $\ell = $ $0^\circ$--$20^\circ$, $20^\circ$--$ 70^\circ$, and $70^\circ $--$180^\circ$. The plots in Figure \ref{fig:distribution_bins2} show that the events at longitudes $\ell < 20^\circ$ are shorter than those at longitudes $\ell > 20^\circ$. This is indicative of tracing events closer to the Galactic bulge, where distances to source and lens are larger and relative proper motions are higher --- both contributing to an overall lower Einstein timescale. Figure \ref{fig:distribution_bins2} suggests that further out in the Galactic plane, there is a difference in the Einstein timescale between bins $\ell = 20^\circ$--$70^\circ$, and $\ell = 70^\circ$--$ 180^\circ$, but the large error bars indicate that we do not have a large enough statistical sample to conclude there is a dependence of Einstein timescale on Galactic longitude for $\ell > 20^\circ$. With a $\sim 4$ times larger sample size of 216 Galactic plane microlensing events, \cite{mroz2020_ogle_plane} also found there was no significant correlation between Einstein timescale and Galactic coordinates within the Galactic plane. We also investigated if there was a statistically significant dependence of Einstein timescale on Galactic latitude and found no significant trend given our current dataset.

\subsection{Comparisons with Theoretical Models and Previous Work}

The only other work to have studied a population of Galactic plane microlensing events has been \cite{mroz2020_ogle_plane}, which found hundreds of microlensing events in the southern Galactic plane. Our results regarding the mean Einstein timescale are in agreement with those of that work, reporting a mean Einstein timescale of $\sim 60$ days, which is on average three times longer than the Einstein timescale of microlensing events toward the Galactic bulge. We find that the normalized number of microlensing events decreases exponentially as a function of Galactic longitude with a characteristic angular scale of $\ell_0 \sim 37^\circ$. \cite{mroz2020_ogle_plane} found the optical depth to microlensing and event rate decrease exponentially with a characteristic angular scale of $\ell_0 \sim 32^\circ$, and ran the simulations of \cite{sajadian2019} to find a predicted angular scale of $\ell_0 \sim 36^\circ$. Our results are consistent with both studies.

\cite{medford2020} calculated microlensing event statistics for ZTF, and predicted $\sim 500$ events would be observed towards the Galactic plane over a three-year-long baseline. We acknowledge disagreement between these this paper and our findings. We explore possible reasons. First, it may be due to deficiencies in our detection algorithm. Second, deficiencies in the model could be possible, since the Galactic model adopted by \cite{medford2020} incorrectly predicts $\langle t_E\rangle$ in the Galactic bulge to be $\sim 10$ days. Observational studies such as \cite{mroz_2019_bulge} find $\langle t_E\rangle \sim 20$ days in the Galactic bulge. Third, the model by \cite{medford2020} may also fail to account for weather conditions and irregular observing gaps that can diminish the amount of usable data acquired by ZTF. 

\section{Conclusion}
We have shown that Galactic plane microlensing events are present in ZTF data and have compiled a list of 60 candidate events, including three that show signatures of the microlensing parallax effect. We presented model parameters for all events, of which the Einstein timescale, $t_\mathrm{E}$, is also used to probe Galactic structure. By injecting simulated microlensing events into ZTF fields where we have found true events in the data, we calculate detection efficiencies which are used to correctly calculate the average timescale of all events: $\langle t_\mathrm{E} \rangle = 61.0 \pm 8.3$ days.

We determine that on the whole, Galactic plane microlensing events ($t_\mathrm{E} \sim 60$ days) are three times longer than events located in the Galactic bulge ($t_\mathrm{E} \sim 20$ days), consistent with prior studies. Furthermore, from our modest sample, we see signatures of Galactic structure, where the normalized number of events decreases exponentially as a function of Galactic longitude with a characteristic angular scale of $\sim37^\circ$. Microlensing events at $\ell < 20^\circ$ have shorter Einstein timescales than those at $\ell > 20^\circ$, demonstrating that we are probing the higher relative motion and larger distance to events at low longitudes closer to the Galactic bulge. A sharp decrease of events at $|b| \approx 0^\circ$ is also indicative of strong reddening at the lowest latitudes of the Galactic plane due to interstellar dust.

Microlensing events taking place in the Galactic plane can now be found within large surveys such as OGLE-GVS, \textit{Gaia}, and ZTF. Through the microlensing parallax effect (both Earth- and space-based), microlensing offer strong prospects for measuring the masses of dark objects in the disk of the Milky Way, in particular isolated objects. 

This study is part of a two-part series. Low- to medium-resolution spectroscopy (R $\sim$ 300 - 5,000) of the 60 candidates will be undertaken and reported. While we believe our detection algorithm has a high level of purity, spectroscopy will filter out the last possible false positives such as Be star outbursts or long-period variable stars. We will determine the distance to our events with low blending ($F_s \gtrsim$ 0.7) and subsequently report the average mass of a dark lens in the Galactic disk from our sample.

\section{Acknowledgements}
The authors thank the ZTF Variable Star Group for insightful discussions. A.C.R acknowledges funding support from the Caltech Anthony Fellowship through the Division of Physics, Mathematics, and Astronomy. 

This work is based on observations obtained with the Samuel Oschin Telescope 48-inch and the 60-inch Telescope at the Palomar Observatory as part of the Zwicky Transient Facility project. Major funding has been provided by the U.S National Science Foundation under Grant No. AST-1440341 and by the ZTF partner institutions: the California Institute of Technology, the Oskar Klein Centre, the Weizmann Institute of Science, the University of Maryland, the University of Washington, Deutsches Elektronen-Synchrotron, the University of Wisconsin-Milwaukee, and the TANGO Program of the University System of Taiwan.

The ZTF forced-photometry service was funded under the Heising-Simons Foundation grant \#12540303 (PI: Graham).

This work has made use of data from the European Space
Agency (ESA) mission Gaia (\url{https://www.cosmos.esa.int/gaia}), processed by the Gaia Data Processing and Analysis Consortium (DPAC, \url{https://www.cosmos.esa.int/web/gaia/ dpac/consortium}). Funding for the DPAC has been provided by national institutions, in particular, the institutions participating in the Gaia Multilateral Agreement.

\bibliography{main}{}
\bibliographystyle{aasjournal}

\appendix
\section{List of All Events}
We present a list of all 60 microlensing events in Table \ref{tab:all_events_ids}, ordered by coordinates, ZTF ID, and Gaia alert (if available) \citep{gaiacollab2013, gaia_ulens2018}. 

In Table \ref{tab:all_events}, we present all model parameters from the MCMC fit including models fitting for parallax (if statistically significant). All errors are reported as the 16th and 84 quartiles of the posterior distribution. $r_s$ and $g_s$ are the source magnitudes in the $r$- and $g$-band, respectively. $f_s$ is the fractional contribution of the source to the total observed flux (i.e. $f_s = F_s/(F_s + F_b)$) where $F_s$ is the source flux and $F_b$ is the blended flux.

\startlongtable
\begin{deluxetable*}{cccc}
\label{tab:all_events_ids}
\tabletypesize{\tiny}
\tablehead{
\colhead{ZTF ID}  & \colhead{RA (J2000)} & \colhead{Decl (J2000)} & \colhead{Gaia Alert} }
\tablecaption{List of Microlensing Events in ZTF Data}
\startdata
ZTF19aabbuqn & 48.694297 & 62.343460 & -  \\
ZTF19aainwvb & 55.197571 & 57.955833 & Gaia19bjq/AT2019dou \\
ZTF19adbsiat & 65.329219 & 30.695007 & -\\
ZTF19aatwaux & 258.208394 & -27.182027 & -\\
ZTF19abpxurg & 260.994091 & -20.194965 & -\\
ZTF20aaukggm & 268.528269 & -20.022328 & -\\
ZTF18ablrdcc & 271.439137 & -12.014536 & Gaia18chq/AT 2018fie\\
ZTF19aawchkq & 271.794669 & -14.315655 & -\\
ZTF19abbwpl & 271.842407 & -15.547389 & -\\
ZTF18ablrbkj & 271.850468 & -10.314384 & -\\
ZTF19aaonska & 273.900546 & -2.256948 & Gaia19awc/AT 2019bpw\\
ZTF20aawyizf \footnote{Strong candidate for binary lensing event upon visual inspection of lightcurve.} & 274.912836 & -10.329083 & -\\
ZTF19aaprbng & 274.913468 & 0.591055 & -\\
ZTF20aauodap & 276.422203 & -6.410752 &  -\\
ZTF20aawanug & 276.484352 & -19.624858 &-\\
ZTF19abhkrjx & 278.163440 & -13.144191 & Gaia19dry/AT 2019ooc\\
ZTF19acigmif & 278.554507 & 3.654422 & - \\
ZTF19aaekacq & 279.404621 & 11.200600 & - \\
ZTF19abibzvr & 279.567874 & -10.709309 & -\\
ZTF18abqbeqv & 279.578747 & 7.837890 &-\\
ZTF20abkymiq & 280.207629 & -10.285631 & -\\
ZTF20abaptby & 280.809384 & -8.669005 & -\\
ZTF19aamrjmu & 280.734524 & 32.873077 & -\\
ZTF19abijroe & 281.724243 & -12.913367 & -\\
ZTF19aavndrc & 281.836916 & -4.338114 & -\\
ZTF20abhmltf & 283.251901 & -19.851325 &-\\
ZTFJ1853.5-0415 & 283.366620 & -4.251845 &-\\
ZTF19aaxsdqz & 283.497181 & -1.152238 &-\\
ZTF19abcpukt & 283.871713 & -10.752056 & Gaia19dur/AT 2019oya \\
ZTFJ1856.2+1007 & 284.049364 & 10.116730 & -\\
ZTF18abhxjmj & 284.029166 & 13.152283 &- \\
ZTF19abgslgl & 284.103651 & -20.323823 &  -\\
ZTF20aawxugf & 284.869773 & -18.107708 &-  \\
ZTFJ1856.2-0110 & 284.061768 & -1.168474 & -\\
ZTF20aawxtfq & 285.568079 & -4.147456 & - \\
ZTF19acctqyc & 285.545821 & -2.919128 &  Gaia19dxg/AT 2019pnt \\
ZTFJ1902.2+0001 & 285.539415 & 0.030082 & Gaia19drz/AT2019ood \\
ZTF18abmoxlq & 285.984024 & -13.929453 & - \\
ZTF19abjtzvc & 286.468609 & -9.741307 &-\\
ZTF19aaoocwc & 287.557242 & 12.559861 & -\\
ZTF18abaqxrt & 290.617205 & 1.706495 & -\\
ZTF19aatudnj & 290.663289 & 19.550419 & Gaia19bzf/AT 2019gdk\\
ZTF18aazdbym & 290.784390 & 7.810464 & -\\
ZTF20abkyuyk & 290.832469 & 23.773103 & - \\
ZTF20aavmhsg & 290.016404 & 37.434801 & - \\
ZTFJ1928.7+3039 & 292.175980 & 30.664609 & -\\
ZTF20abmxjsq & 293.061917 & 25.482553 & -\\
ZTF20abrtvbz & 294.844574 & 39.178278 &-\\
ZTF19aavisrq & 297.706150 & 34.637374 & Gaia19brt/AT 2019etl \\
ZTF19acaagdx & 298.683899 & 5.402632 & -\\
ZTF20abohkdo & 298.050406 & 26.993506 & -\\
ZTF18abtopdh & 299.292605 & 35.443629 & -\\
ZTF20abbynqb & 301.318168 & 55.322594 & - \\
ZTF20abxwenr & 301.549797 & 35.631697 & Gaia20eha/AT 2020tbt \\
ZTF18absrqlr & 307.149384 & 22.830472 & Gaia18cmk/AT 2018fug \\
ZTF19aavnrqt & 309.034117 & 32.720917 & Gaia19dae/AT 2019lje \\
ZTF19abftuld & 318.263352 & 43.337654 & Gaia19cyv/AT 2019lhm \\
ZTF20abvwhlb & 324.577345 & 48.479250 & Gaia20ebu/AT 2020sjy  \\
ZTF18aaztjyd & 326.173126 & 59.377905 & - \\
ZTF18aayhjoe & 329.192958 & 54.098562 &-\\
\enddata
\end{deluxetable*}

\begin{longrotatetable}
\begin{deluxetable*}{cccccccccccccc}
\label{tab:all_events}
\tabletypesize{\tiny}
\tablehead{
\colhead{ID}  & \colhead{$t_0 \textrm{ (HJD')}$} & \colhead{$t_\mathrm{E} \textrm{ (days)}$} & \colhead{$u_0$} &
\colhead{$\pi_\textrm{EN}$} & \colhead{$\pi_\textrm{EE}$} & \colhead{$r_s \textrm{ (mag)}$} & \colhead{$f_{s,r}$} & \colhead{$g_s \textrm{ (mag)}$} & \colhead{$f_{s,g}$} & \colhead{$\chi^2$/d.o.f.} & \colhead{}
}
\tablecaption{List of Microlensing Events in ZTF Data with MCMC Modeling Parameters}
\startdata
ZTF19aabbuqn  & $8506.17^{+0.39}_{-0.37}$ & $38.62^{+8.85}_{-7.68}$ & $0.281^{+0.106}_{-0.070}$ & - & - & $19.85^{+0.38}_{-0.46}$ & $0.26^{+0.10}_{-0.07}$ & $21.30^{+0.38}_{-0.46}$ & $0.28^{+0.11}_{-0.07}$ & 831.4/413 & \\
ZTF19aainwvb  & $8658.05^{+0.45}_{-0.44}$ & $160.12^{+2.41}_{-2.37}$ & $0.444^{+0.011}_{-0.011}$ & - & - & $18.28^{+0.04}_{-0.04}$ & $1.30^{+0.06}_{-0.06}$ & $19.39^{+0.04}_{-0.04}$ & $1.31^{+0.06}_{-0.06}$ & 7819.5/2189 & \\
ZTF19aainwvb  & $8625.27^{+0.52}_{-0.50}$ & $185.92^{+3.78}_{-3.91}$ & $0.128^{+0.005}_{-0.005}$ & $-0.29^{+0.01}_{-0.01}$ & $0.00^{+0.00}_{-0.00}$ & $19.02^{+0.03}_{-0.04}$ & $0.64^{+0.01}_{-0.02}$ & $20.10^{+0.03}_{-0.04}$ & $0.67^{+0.02}_{-0.02}$ & 4584.6/2189 & \\
ZTF19aainwvb  & $8618.20^{+0.68}_{-0.66}$ & $333.45^{+20.21}_{-17.58}$ & $0.025^{+0.006}_{-0.006}$ & $-0.17^{+0.01}_{-0.01}$ & $-0.04^{+0.00}_{-0.00}$ & $19.76^{+0.07}_{-0.07}$ & $0.33^{+0.02}_{-0.02}$ & $20.85^{+0.07}_{-0.07}$ & $0.34^{+0.02}_{-0.02}$ & 4452.5/2189 & \\
ZTF19aainwvb  & $8618.27^{+0.75}_{-0.75}$ & $333.35^{+20.36}_{-17.97}$ & $-0.025^{+0.007}_{-0.007}$ & $0.17^{+0.01}_{-0.01}$ & $0.04^{+0.00}_{-0.00}$ & $19.76^{+0.07}_{-0.07}$ & $0.33^{+0.02}_{-0.02}$ & $20.85^{+0.07}_{-0.07}$ & $0.34^{+0.02}_{-0.02}$ & 4452.5/2189 & \\
ZTF19aainwvb  & $8625.29^{+0.51}_{-0.49}$ & $185.85^{+3.73}_{-3.76}$ & $-0.129^{+0.005}_{-0.005}$ & $0.29^{+0.01}_{-0.01}$ & $-0.00^{+0.00}_{-0.00}$ & $19.02^{+0.03}_{-0.03}$ & $0.64^{+0.01}_{-0.01}$ & $20.10^{+0.03}_{-0.03}$ & $0.67^{+0.01}_{-0.02}$ & 4584.6/2189 & \\
ZTF19adbsiat  & $8915.50^{+0.81}_{-0.81}$ & $58.86^{+6.47}_{-4.85}$ & $0.573^{+0.091}_{-0.088}$ & - & - & $18.56^{+0.27}_{-0.26}$ & $0.73^{+0.13}_{-0.15}$ & $19.57^{+0.27}_{-0.25}$ & $0.72^{+0.13}_{-0.14}$ & 792.6/908 & \\
ZTF19aatwaux  & $8637.05^{+0.35}_{-0.34}$ & $50.25^{+4.01}_{-2.90}$ & $0.179^{+0.021}_{-0.024}$ & - & - & $19.13^{+0.15}_{-0.12}$ & $0.93^{+0.11}_{-0.10}$ & $20.50^{+0.15}_{-0.12}$ & $0.88^{+0.10}_{-0.09}$ & 1155.7/432 & \\
ZTF19abpxurg  & $8713.62^{+0.04}_{-0.04}$ & $25.86^{+1.39}_{-1.24}$ & $0.062^{+0.006}_{-0.006}$ & - & - & $19.34^{+0.09}_{-0.08}$ & $0.98^{+0.07}_{-0.07}$ & $20.55^{+0.09}_{-0.08}$ & $0.85^{+0.06}_{-0.06}$ & 734.2/732 & \\
ZTF20aaukggm & $8956.33^{+0.30}_{-0.28}$ & $54.23^{+3.22}_{-2.86}$ & $0.094^{+0.014}_{-0.014}$ & - & - & $17.83^{+0.10}_{-0.10}$ & $0.88^{+0.07}_{-0.07}$ & $20.24^{+0.10}_{-0.10}$ & $0.77^{+0.06}_{-0.06}$ & 1806.3/459 & \\
ZTF18ablrdcc  & $8353.81^{+0.34}_{-0.35}$ & $56.56^{+8.15}_{-6.15}$ & $0.150^{+0.030}_{-0.030}$ & - & - & $20.24^{+0.22}_{-0.20}$ & $0.99^{+0.18}_{-0.20}$ & - & - & 385.1/435 & \\
ZTF19aawchkq  & $8633.36^{+1.03}_{-1.02}$ & $55.42^{+19.12}_{-13.82}$ & $0.198^{+0.108}_{-0.069}$ & - & - & $20.73^{+0.52}_{-0.56}$ & $0.14^{+0.08}_{-0.05}$ & - & - & 405.2/398 & \\
ZTF19abbwpl  & $8677.90^{+0.20}_{-0.20}$ & $30.42^{+3.61}_{-3.62}$ & $0.530^{+0.116}_{-0.084}$ & - & - & $18.02^{+0.27}_{-0.34}$ & $0.33^{+0.08}_{-0.06}$ & $20.17^{+0.28}_{-0.35}$ & $0.30^{+0.08}_{-0.06}$ & 985.4/473 & \\
ZTF18ablrbkj  & $8262.27^{+0.92}_{-0.98}$ & $104.45^{+125.13}_{-42.82}$ & $0.121^{+0.114}_{-0.072}$ & - & - & $21.82^{+1.03}_{-0.83}$ & $0.40^{+0.29}_{-0.23}$ & - & - & 179.6/322 & \\
ZTF19aaonska  & $8612.70^{+0.23}_{-0.22}$ & $68.74^{+2.65}_{-2.54}$ & $0.276^{+0.017}_{-0.017}$ & - & - & $19.09^{+0.08}_{-0.08}$ & $0.91^{+0.06}_{-0.06}$ & $22.13^{+0.08}_{-0.08}$ & $0.49^{+0.03}_{-0.03}$ & 627.9/647 & \\
ZTF20aawyizf & $9004.36^{+1.52}_{-1.72}$ & $1024.09^{+2375.70}_{-525.33}$ & $0.029^{+0.032}_{-0.021}$ & - & - & $23.14^{+1.33}_{-0.81}$ & $0.02^{+0.02}_{-0.01}$ & - & - & 302.3/469 & \\
ZTF19aaprbng  & $8629.03^{+1.30}_{-1.28}$ & $209.78^{+61.52}_{-44.84}$ & $0.291^{+0.105}_{-0.077}$ & - & - & $20.16^{+0.41}_{-0.44}$ & $0.22^{+0.09}_{-0.06}$ & $21.55^{+0.41}_{-0.44}$ & $0.23^{+0.09}_{-0.06}$ & 735.5/788 & \\
ZTF20aauodap  & $8976.89^{+0.38}_{-0.35}$ & $69.77^{+8.49}_{-5.65}$ & $0.158^{+0.020}_{-0.023}$ & - & - & $19.69^{+0.19}_{-0.14}$ & $1.01^{+0.16}_{-0.14}$ & - & - & 256.0/459 & \\
ZTF20aawanug  & $8980.72^{+0.05}_{-0.05}$ & $54.16^{+3.41}_{-3.09}$ & $0.113^{+0.010}_{-0.009}$ & - & - & $18.09^{+0.09}_{-0.09}$ & $0.51^{+0.03}_{-0.03}$ & $19.64^{+0.09}_{-0.09}$ & $0.52^{+0.03}_{-0.03}$ & 1378.6/738 & \\
ZTF19abhkrjx  & $8699.74^{+0.01}_{-0.01}$ & $22.71^{+0.45}_{-0.42}$ & $0.120^{+0.003}_{-0.003}$ & - & - & $18.24^{+0.03}_{-0.03}$ & $1.15^{+0.03}_{-0.03}$ & $20.88^{+0.03}_{-0.03}$ & $0.15^{+0.00}_{-0.00}$ & 2282.3/912 & \\
ZTF19acigmif  & $8793.48^{+0.67}_{-0.69}$ & $44.70^{+17.04}_{-14.11}$ & $0.195^{+0.144}_{-0.070}$ & - & - & $19.54^{+0.54}_{-0.75}$ & $0.15^{+0.12}_{-0.05}$ & $21.27^{+0.53}_{-0.73}$ & $0.16^{+0.13}_{-0.06}$ & 1296.2/965 & \\
ZTF19aaekacq  & $8546.02^{+0.57}_{-0.56}$ & $69.35^{+7.14}_{-7.22}$ & $0.428^{+0.117}_{-0.085}$ & - & - & $18.10^{+0.27}_{-0.33}$ & $0.36^{+0.09}_{-0.07}$ & $19.04^{+0.26}_{-0.32}$ & $0.34^{+0.08}_{-0.06}$ & 1382.3/846 & \\
ZTF19abibzvr  & $8300.95^{+0.40}_{-0.39}$ & $30.27^{+6.61}_{-3.45}$ & $0.331^{+0.066}_{-0.085}$ & - & - & $20.26^{+0.40}_{-0.25}$ & $0.98^{+0.30}_{-0.26}$ & - & - & 499.2/809 & \\
ZTF18abqbeqv  & $8386.97^{+1.33}_{-1.29}$ & $68.68^{+16.96}_{-12.30}$ & $0.690^{+0.244}_{-0.197}$ & - & - & $17.80^{+0.59}_{-0.61}$ & $0.35^{+0.18}_{-0.14}$ & $18.81^{+0.59}_{-0.61}$ & $0.39^{+0.19}_{-0.15}$ & 1928.3/975 & \\
ZTF20abkymiq  & $9088.46^{+0.06}_{-0.06}$ & $83.77^{+1.58}_{-1.52}$ & $0.077^{+0.002}_{-0.002}$ & - & - & $18.57^{+0.03}_{-0.03}$ & $0.81^{+0.02}_{-0.02}$ & $21.38^{+0.03}_{-0.03}$ & $0.96^{+0.02}_{-0.02}$ & 1033.5/914 & \\
ZTF20abaptby  & $8991.77^{+0.11}_{-0.11}$ & $12.69^{+0.95}_{-0.71}$ & $0.261^{+0.028}_{-0.030}$ & - & - & $18.47^{+0.16}_{-0.13}$ & $0.86^{+0.10}_{-0.10}$ & - & - & 631.1/813 & \\
ZTF19aamrjmu  & $8579.63^{+0.11}_{-0.11}$ & $66.07^{+3.09}_{-2.92}$ & $0.133^{+0.009}_{-0.009}$ & - & - & $20.15^{+0.07}_{-0.08}$ & $0.74^{+0.04}_{-0.04}$ & $21.07^{+0.07}_{-0.08}$ & $0.84^{+0.05}_{-0.05}$ & 1623.5/1722 & \\
ZTF19abijroe  & $8695.88^{+0.07}_{-0.07}$ & $20.81^{+2.03}_{-1.77}$ & $0.284^{+0.034}_{-0.032}$ & - & - & $20.17^{+0.16}_{-0.16}$ & $1.58^{+0.37}_{-0.42}$ & - & - & 644.3/719 & \\
ZTF19aavndrc  & $8637.35^{+0.08}_{-0.08}$ & $74.89^{+3.12}_{-2.91}$ & $0.091^{+0.005}_{-0.005}$ & - & - & $19.96^{+0.07}_{-0.06}$ & $0.38^{+0.02}_{-0.02}$ & - & - & 901.2/799 & \\
ZTF20abhmltf  & $9044.66^{+0.44}_{-0.44}$ & $40.98^{+3.59}_{-2.08}$ & $0.404^{+0.034}_{-0.049}$ & - & - & $18.31^{+0.19}_{-0.13}$ & $0.89^{+0.13}_{-0.10}$ & $19.08^{+0.19}_{-0.12}$ & $0.87^{+0.12}_{-0.09}$ & 841.8/700 & \\
ZTFJ1853.5-0415  & $8291.99^{+0.52}_{-0.52}$ & $11.87^{+0.64}_{-0.58}$ & $0.811^{+0.033}_{-0.055}$ & - & - & $17.89^{+0.14}_{-0.08}$ & $0.94^{+0.11}_{-0.07}$ & - & - & 1322.7/806 & \\
ZTF19aaxsdqz  & $8676.11^{+0.06}_{-0.06}$ & $61.87^{+2.20}_{-1.97}$ & $0.177^{+0.008}_{-0.008}$ & - & - & $19.01^{+0.06}_{-0.05}$ & $1.03^{+0.05}_{-0.05}$ & $21.73^{+0.06}_{-0.05}$ & $0.74^{+0.03}_{-0.03}$ & 730.7/674 & \\
ZTF19abcpukt  & $8691.38^{+0.00}_{-0.00}$ & $446.70^{+141.70}_{-74.89}$ & $0.003^{+0.001}_{-0.001}$ & - & - & $21.51^{+0.30}_{-0.20}$ & $0.38^{+0.05}_{-0.06}$ & - & - & 5193.3/791 & \\
ZTFJ1856.2+1007  & $8367.43^{+0.05}_{-0.05}$ & $29.79^{+4.80}_{-4.01}$ & $0.051^{+0.010}_{-0.008}$ & - & - & $21.79^{+0.20}_{-0.19}$ & $0.16^{+0.03}_{-0.02}$ & $22.88^{+0.20}_{-0.19}$ & $0.19^{+0.03}_{-0.03}$ & 777.3/932 & \\
ZTF18abhxjmj  & $8249.16^{+0.22}_{-0.23}$ & $35.35^{+1.31}_{-1.21}$ & $0.256^{+0.015}_{-0.015}$ & - & - & $19.26^{+0.08}_{-0.08}$ & $1.64^{+0.20}_{-0.21}$ & $20.69^{+0.07}_{-0.07}$ & $1.62^{+0.19}_{-0.20}$ & 703.0/903 & \\
ZTF19abgslgl  & $8683.43^{+0.37}_{-0.35}$ & $37.33^{+6.17}_{-4.97}$ & $0.159^{+0.044}_{-0.035}$ & - & - & $21.08^{+0.27}_{-0.27}$ & $0.35^{+0.07}_{-0.06}$ & $21.73^{+0.25}_{-0.26}$ & $0.39^{+0.07}_{-0.06}$ & 743.0/666 & \\
ZTF20aawxugf  & $8969.91^{+1.08}_{-1.14}$ & $49.30^{+5.92}_{-3.82}$ & $0.500^{+0.061}_{-0.083}$ & - & - & $18.70^{+0.29}_{-0.19}$ & $0.78^{+0.15}_{-0.12}$ & $19.42^{+0.29}_{-0.19}$ & $0.87^{+0.18}_{-0.15}$ & 822.3/718 & \\
ZTFJ1856.2-0110  & $8352.71^{+0.26}_{-0.25}$ & $20.20^{+11.68}_{-6.54}$ & $0.137^{+0.124}_{-0.069}$ & - & - & $19.98^{+0.71}_{-0.72}$ & $0.19^{+0.15}_{-0.08}$ & $21.27^{+0.73}_{-0.73}$ & $0.25^{+0.18}_{-0.11}$ & 222.3/994 & \\
ZTF20aawxtfq  & $8975.61^{+0.62}_{-0.66}$ & $76.98^{+36.21}_{-16.61}$ & $0.096^{+0.042}_{-0.038}$ & - & - & $21.41^{+0.57}_{-0.41}$ & $0.04^{+0.02}_{-0.02}$ & $23.20^{+0.57}_{-0.40}$ & $0.02^{+0.01}_{-0.01}$ & 1318.7/1035 & \\
ZTF19acctqyc  & $8756.40^{+0.98}_{-0.93}$ & $39.21^{+2.47}_{-1.99}$ & $0.470^{+0.036}_{-0.044}$ & - & - & $19.47^{+0.16}_{-0.12}$ & $1.07^{+0.15}_{-0.14}$ & $20.84^{+0.15}_{-0.12}$ & $1.11^{+0.16}_{-0.14}$ & 783.9/937 & \\
Gaia19drz/AT2019ood  & $8750.29^{+0.95}_{-0.93}$ & $83.36^{+13.11}_{-11.20}$ & $0.331^{+0.089}_{-0.068}$ & - & - & $19.76^{+0.31}_{-0.34}$ & $0.50^{+0.11}_{-0.11}$ & $21.22^{+0.32}_{-0.35}$ & $0.60^{+0.13}_{-0.15}$ & 623.6/930 & \\
ZTF18abmoxlq  & $8322.09^{+0.36}_{-0.35}$ & $111.34^{+28.77}_{-20.33}$ & $0.168^{+0.049}_{-0.041}$ & - & - & $19.18^{+0.35}_{-0.33}$ & $0.26^{+0.07}_{-0.06}$ & $20.02^{+0.35}_{-0.33}$ & $0.28^{+0.07}_{-0.06}$ & 1039.3/721 & \\
ZTF19abjtzvc  & $8709.79^{+0.38}_{-0.39}$ & $98.32^{+35.35}_{-22.21}$ & $0.218^{+0.086}_{-0.068}$ & - & - & $20.63^{+0.48}_{-0.45}$ & $0.40^{+0.14}_{-0.12}$ & - & - & 575.5/594 & \\
ZTF19aaoocwc  & $8579.41^{+0.53}_{-0.61}$ & $52.02^{+7.04}_{-5.22}$ & $0.226^{+0.047}_{-0.045}$ & - & - & $19.97^{+0.28}_{-0.25}$ & $0.83^{+0.16}_{-0.18}$ & - & - & 577.1/720 & \\
ZTF18abaqxrt  & $8302.65^{+0.21}_{-0.20}$ & $33.44^{+3.27}_{-2.29}$ & $0.606^{+0.072}_{-0.080}$ & - & - & $16.34^{+0.24}_{-0.20}$ & $0.76^{+0.12}_{-0.12}$ & - & - & 863.0/256 & \\
ZTF19aatudnj  & $8636.82^{+0.47}_{-0.45}$ & $615.12^{+869.35}_{-206.47}$ & $0.027^{+0.015}_{-0.016}$ & - & - & $22.73^{+0.98}_{-0.47}$ & $0.02^{+0.01}_{-0.01}$ & $24.53^{+0.98}_{-0.47}$ & $0.02^{+0.01}_{-0.01}$ & 1905.9/1169 & \\
ZTF18aazdbym  & $8272.97^{+0.25}_{-0.26}$ & $20.28^{+2.89}_{-2.08}$ & $0.488^{+0.093}_{-0.095}$ & - & - & $18.15^{+0.34}_{-0.30}$ & $0.63^{+0.13}_{-0.14}$ & $18.99^{+0.33}_{-0.29}$ & $0.70^{+0.14}_{-0.16}$ & 896.9/979 & \\
ZTF20abkyuyk  & $9085.44^{+0.07}_{-0.07}$ & $49.56^{+1.84}_{-1.75}$ & $0.261^{+0.014}_{-0.014}$ & - & - & $18.38^{+0.07}_{-0.07}$ & $0.45^{+0.02}_{-0.02}$ & $19.76^{+0.07}_{-0.07}$ & $0.44^{+0.02}_{-0.02}$ & 2420.3/1245 & \\
ZTF20aavmhsg  & $8983.14^{+0.09}_{-0.09}$ & $46.43^{+3.01}_{-2.78}$ & $0.120^{+0.011}_{-0.010}$ & - & - & $19.74^{+0.10}_{-0.10}$ & $0.36^{+0.03}_{-0.02}$ & $20.90^{+0.10}_{-0.10}$ & $0.35^{+0.02}_{-0.02}$ & 1022.8/1153 & \\
ZTFJ1928.7+3039  & $8369.91^{+0.20}_{-0.21}$ & $50.10^{+11.62}_{-8.27}$ & $0.064^{+0.018}_{-0.015}$ & - & - & $22.07^{+0.31}_{-0.28}$ & $0.49^{+0.09}_{-0.09}$ & $23.48^{+0.30}_{-0.27}$ & $0.15^{+0.04}_{-0.03}$ & 477.7/589 & \\
ZTF20abmxjsq  & $9051.45^{+0.52}_{-0.51}$ & $115.51^{+56.04}_{-28.49}$ & $0.139^{+0.061}_{-0.053}$ & - & - & $21.64^{+0.57}_{-0.45}$ & $0.09^{+0.04}_{-0.04}$ & $23.37^{+0.57}_{-0.45}$ & $0.07^{+0.04}_{-0.03}$ & 931.3/1120 & \\
ZTF20abrtvbz  & $9088.92^{+0.22}_{-0.22}$ & $59.90^{+9.09}_{-7.62}$ & $0.117^{+0.024}_{-0.020}$ & - & - & $21.35^{+0.21}_{-0.22}$ & $0.17^{+0.03}_{-0.03}$ & $22.12^{+0.21}_{-0.22}$ & $0.15^{+0.03}_{-0.02}$ & 1243.2/1146 & \\
ZTF19aavisrq  & $8651.81^{+0.01}_{-0.01}$ & $104.92^{+4.62}_{-4.33}$ & $0.015^{+0.001}_{-0.001}$ & - & - & $20.69^{+0.05}_{-0.05}$ & $0.14^{+0.01}_{-0.01}$ & $21.71^{+0.05}_{-0.05}$ & $0.11^{+0.00}_{-0.00}$ & 1446.4/1243 & \\
ZTF19acaagdx  & $8762.59^{+0.04}_{-0.04}$ & $115.99^{+9.21}_{-8.00}$ & $0.010^{+0.001}_{-0.001}$ & - & - & $21.83^{+0.10}_{-0.09}$ & $0.11^{+0.01}_{-0.01}$ & $22.49^{+0.09}_{-0.08}$ & $0.12^{+0.01}_{-0.01}$ & 1829.6/1461 & \\
ZTF20abohkdo  & $9061.86^{+0.27}_{-0.28}$ & $29.66^{+8.17}_{-6.39}$ & $0.174^{+0.076}_{-0.048}$ & - & - & $21.07^{+0.39}_{-0.44}$ & $0.14^{+0.06}_{-0.04}$ & - & - & 597.1/809 & \\
ZTF18abtopdh  & $8367.88^{+0.03}_{-0.03}$ & $24.67^{+1.76}_{-1.69}$ & $0.062^{+0.007}_{-0.006}$ & - & - & $20.92^{+0.10}_{-0.10}$ & $0.34^{+0.02}_{-0.02}$ & $22.56^{+0.10}_{-0.11}$ & $0.27^{+0.02}_{-0.02}$ & 852.3/1005 & \\
ZTF20abbynqb  & $9045.89^{+0.17}_{-0.17}$ & $50.97^{+3.34}_{-3.39}$ & $0.385^{+0.043}_{-0.035}$ & - & - & $17.45^{+0.14}_{-0.16}$ & $0.39^{+0.04}_{-0.04}$ & $18.53^{+0.14}_{-0.16}$ & $0.36^{+0.04}_{-0.03}$ & 1461.2/1149 & \\
ZTF20abbynqb  & $9045.19^{+0.17}_{-0.18}$ & $31.58^{+1.89}_{-1.11}$ & $0.658^{+0.034}_{-0.051}$ & $-0.97^{+0.15}_{-0.12}$ & $-0.48^{+0.05}_{-0.05}$ & $16.55^{+0.15}_{-0.09}$ & $0.90^{+0.10}_{-0.07}$ & $17.62^{+0.15}_{-0.09}$ & $0.83^{+0.09}_{-0.06}$ & 1379.2/1149 & \\
ZTF20abbynqb  & $9045.20^{+0.17}_{-0.18}$ & $31.75^{+1.89}_{-1.23}$ & $-0.655^{+0.053}_{-0.036}$ & $0.96^{+0.12}_{-0.16}$ & $0.48^{+0.05}_{-0.05}$ & $16.57^{+0.15}_{-0.10}$ & $0.89^{+0.10}_{-0.08}$ & $17.63^{+0.15}_{-0.10}$ & $0.82^{+0.09}_{-0.07}$ & 1379.3/1149 & \\
ZTF20abxwenr  & $9117.33^{+0.01}_{-0.01}$ & $162.56^{+18.47}_{-14.23}$ & $0.006^{+0.001}_{-0.001}$ & - & - & $22.05^{+0.12}_{-0.10}$ & $0.31^{+0.02}_{-0.02}$ & - & - & 643.4/655 & \\
ZTF18absrqlr & $8362.83^{+0.12}_{-0.11}$ & $37.17^{+2.30}_{-2.06}$ & $0.310^{+0.028}_{-0.027}$ & - & - & $18.88^{+0.12}_{-0.12}$ & $0.66^{+0.05}_{-0.06}$ & $19.56^{+0.12}_{-0.12}$ & $0.72^{+0.06}_{-0.06}$ & 1895.3/1380 & \\
ZTF19aavnrqt  & $8722.08^{+0.20}_{-0.19}$ & $78.26^{+0.84}_{-0.63}$ & $0.642^{+0.006}_{-0.009}$ & - & - & $17.10^{+0.03}_{-0.02}$ & $1.00^{+0.02}_{-0.02}$ & $19.07^{+0.03}_{-0.02}$ & $0.72^{+0.01}_{-0.01}$ & 2305.6/1314 & \\
ZTF19aavnrqt  & $8722.32^{+0.57}_{-0.42}$ & $140.18^{+125.15}_{-41.93}$ & $0.431^{+0.165}_{-0.171}$ & $0.17^{+0.04}_{-0.04}$ & $0.23^{+0.04}_{-0.05}$ & $17.79^{+0.75}_{-0.55}$ & $0.53^{+0.21}_{-0.22}$ & $19.76^{+0.75}_{-0.55}$ & $0.38^{+0.17}_{-0.15}$ & 2277.0/1314 & \\
ZTF19aavnrqt  & $8722.56^{+1.26}_{-0.76}$ & $284.97^{+382.15}_{-169.15}$ & $-0.257^{+0.105}_{-0.259}$ & $-0.21^{+0.06}_{-0.07}$ & $-0.23^{+0.07}_{-0.05}$ & $18.55^{+0.68}_{-1.06}$ & $0.26^{+0.32}_{-0.15}$ & $20.52^{+0.68}_{-1.06}$ & $0.19^{+0.25}_{-0.09}$ & 2269.7/1314 & \\
ZTF19abftuld  & $8727.95^{+0.12}_{-0.11}$ & $255.46^{+79.05}_{-45.10}$ & $0.072^{+0.017}_{-0.018}$ & - & - & $21.46^{+0.32}_{-0.24}$ & $0.14^{+0.03}_{-0.03}$ & $22.59^{+0.33}_{-0.24}$ & $0.14^{+0.03}_{-0.03}$ & 3405.5/1295 & \\
ZTF20abvwhlb  & $9090.35^{+0.02}_{-0.02}$ & $8.10^{+0.28}_{-0.24}$ & $0.247^{+0.010}_{-0.012}$ & - & - & $18.70^{+0.06}_{-0.05}$ & $1.05^{+0.06}_{-0.06}$ & $19.55^{+0.06}_{-0.05}$ & $1.04^{+0.06}_{-0.06}$ & 1807.0/1886 & \\
ZTF18aaztjyd & $8289.74^{+0.18}_{-0.17}$ & $86.46^{+11.64}_{-8.97}$ & $0.101^{+0.015}_{-0.015}$ & - & - & $21.33^{+0.18}_{-0.16}$ & $0.66^{+0.07}_{-0.08}$ & $22.67^{+0.18}_{-0.16}$ & $0.47^{+0.05}_{-0.05}$ & 574.3/718 & \\
ZTF18aayhjoe  & $8279.00^{+0.23}_{-0.24}$ & $25.94^{+1.52}_{-1.29}$ & $0.536^{+0.044}_{-0.045}$ & - & - & $19.62^{+0.14}_{-0.13}$ & $1.15^{+0.16}_{-0.17}$ & $21.12^{+0.14}_{-0.13}$ & $1.04^{+0.13}_{-0.14}$ & 2006.4/1985 & \\
\enddata
\end{deluxetable*}
\end{longrotatetable}

\end{document}